\documentclass[aps,prd,showpacs,letterpaper,twocolumn,longbibliography,superscriptaddress]{revtex4-1}
\usepackage{graphicx}
\usepackage{bm}
\usepackage{epsf}
\usepackage{rotating}
\usepackage{epsfig,graphics,rotate,color}
\usepackage{wrapfig}
\usepackage{amssymb}
\usepackage{amsmath}
\usepackage{amsfonts}
\usepackage{array,hhline,dcolumn}%
\usepackage{gensymb}
\usepackage{placeins}
\usepackage{tabularx}
\usepackage[colorlinks=true,citecolor=blue,linkcolor=blue]{hyperref}

\begin{document}

\title{Search for Nonstandard Neutrino Interactions with IceCube DeepCore}

\affiliation{III. Physikalisches Institut, RWTH Aachen University, D-52056 Aachen, Germany}
\affiliation{Department of Physics, University of Adelaide, Adelaide, 5005, Australia}
\affiliation{Dept.~of Physics and Astronomy, University of Alaska Anchorage, 3211 Providence Dr., Anchorage, AK 99508, USA}
\affiliation{Dept.~of Physics, University of Texas at Arlington, 502 Yates St., Science Hall Rm 108, Box 19059, Arlington, TX 76019, USA}
\affiliation{CTSPS, Clark-Atlanta University, Atlanta, GA 30314, USA}
\affiliation{School of Physics and Center for Relativistic Astrophysics, Georgia Institute of Technology, Atlanta, GA 30332, USA}
\affiliation{Dept.~of Physics, Southern University, Baton Rouge, LA 70813, USA}
\affiliation{Dept.~of Physics, University of California, Berkeley, CA 94720, USA}
\affiliation{Lawrence Berkeley National Laboratory, Berkeley, CA 94720, USA}
\affiliation{Institut f\"ur Physik, Humboldt-Universit\"at zu Berlin, D-12489 Berlin, Germany}
\affiliation{Fakult\"at f\"ur Physik \& Astronomie, Ruhr-Universit\"at Bochum, D-44780 Bochum, Germany}
\affiliation{Universit\'e Libre de Bruxelles, Science Faculty CP230, B-1050 Brussels, Belgium}
\affiliation{Vrije Universiteit Brussel (VUB), Dienst ELEM, B-1050 Brussels, Belgium}
\affiliation{Dept.~of Physics, Massachusetts Institute of Technology, Cambridge, MA 02139, USA}
\affiliation{Dept. of Physics and Institute for Global Prominent Research, Chiba University, Chiba 263-8522, Japan}
\affiliation{Dept.~of Physics and Astronomy, University of Canterbury, Private Bag 4800, Christchurch, New Zealand}
\affiliation{Dept.~of Physics, University of Maryland, College Park, MD 20742, USA}
\affiliation{Dept.~of Physics and Center for Cosmology and Astro-Particle Physics, Ohio State University, Columbus, OH 43210, USA}
\affiliation{Dept.~of Astronomy, Ohio State University, Columbus, OH 43210, USA}
\affiliation{Niels Bohr Institute, University of Copenhagen, DK-2100 Copenhagen, Denmark}
\affiliation{Dept.~of Physics, TU Dortmund University, D-44221 Dortmund, Germany}
\affiliation{Dept.~of Physics and Astronomy, Michigan State University, East Lansing, MI 48824, USA}
\affiliation{Dept.~of Physics, University of Alberta, Edmonton, Alberta, Canada T6G 2E1}
\affiliation{Erlangen Centre for Astroparticle Physics, Friedrich-Alexander-Universit\"at Erlangen-N\"urnberg, D-91058 Erlangen, Germany}
\affiliation{D\'epartement de physique nucl\'eaire et corpusculaire, Universit\'e de Gen\`eve, CH-1211 Gen\`eve, Switzerland}
\affiliation{Dept.~of Physics and Astronomy, University of Gent, B-9000 Gent, Belgium}
\affiliation{Dept.~of Physics and Astronomy, University of California, Irvine, CA 92697, USA}
\affiliation{Dept.~of Physics and Astronomy, University of Kansas, Lawrence, KS 66045, USA}
\affiliation{SNOLAB, 1039 Regional Road 24, Creighton Mine 9, Lively, ON, Canada P3Y 1N2}
\affiliation{Dept.~of Astronomy, University of Wisconsin, Madison, WI 53706, USA}
\affiliation{Dept.~of Physics and Wisconsin IceCube Particle Astrophysics Center, University of Wisconsin, Madison, WI 53706, USA}
\affiliation{Institute of Physics, University of Mainz, Staudinger Weg 7, D-55099 Mainz, Germany}
\affiliation{Department of Physics, Marquette University, Milwaukee, WI, 53201, USA}
\affiliation{Universit\'e de Mons, 7000 Mons, Belgium}
\affiliation{Physik-department, Technische Universit\"at M\"unchen, D-85748 Garching, Germany}
\affiliation{Institut f\"ur Kernphysik, Westf\"alische Wilhelms-Universit\"at M\"unster, D-48149 M\"unster, Germany}
\affiliation{Bartol Research Institute and Dept.~of Physics and Astronomy, University of Delaware, Newark, DE 19716, USA}
\affiliation{Dept.~of Physics, Yale University, New Haven, CT 06520, USA}
\affiliation{Dept.~of Physics, University of Oxford, 1 Keble Road, Oxford OX1 3NP, UK}
\affiliation{Dept.~of Physics, Drexel University, 3141 Chestnut Street, Philadelphia, PA 19104, USA}
\affiliation{Physics Department, South Dakota School of Mines and Technology, Rapid City, SD 57701, USA}
\affiliation{Dept.~of Physics, University of Wisconsin, River Falls, WI 54022, USA}
\affiliation{Dept.~of Physics and Astronomy, University of Rochester, Rochester, NY 14627, USA}
\affiliation{Oskar Klein Centre and Dept.~of Physics, Stockholm University, SE-10691 Stockholm, Sweden}
\affiliation{Dept.~of Physics and Astronomy, Stony Brook University, Stony Brook, NY 11794-3800, USA}
\affiliation{Dept.~of Physics, Sungkyunkwan University, Suwon 440-746, Korea}
\affiliation{Dept.~of Physics and Astronomy, University of Alabama, Tuscaloosa, AL 35487, USA}
\affiliation{Dept.~of Astronomy and Astrophysics, Pennsylvania State University, University Park, PA 16802, USA}
\affiliation{Dept.~of Physics, Pennsylvania State University, University Park, PA 16802, USA}
\affiliation{Dept.~of Physics and Astronomy, Uppsala University, Box 516, S-75120 Uppsala, Sweden}
\affiliation{Dept.~of Physics, University of Wuppertal, D-42119 Wuppertal, Germany}
\affiliation{DESY, D-15738 Zeuthen, Germany}

\author{M.~G.~Aartsen}
\affiliation{Department of Physics, University of Adelaide, Adelaide, 5005, Australia}
\author{M.~Ackermann}
\affiliation{DESY, D-15738 Zeuthen, Germany}
\author{J.~Adams}
\affiliation{Dept.~of Physics and Astronomy, University of Canterbury, Private Bag 4800, Christchurch, New Zealand}
\author{J.~A.~Aguilar}
\affiliation{Universit\'e Libre de Bruxelles, Science Faculty CP230, B-1050 Brussels, Belgium}
\author{M.~Ahlers}
\affiliation{Niels Bohr Institute, University of Copenhagen, DK-2100 Copenhagen, Denmark}
\author{M.~Ahrens}
\affiliation{Oskar Klein Centre and Dept.~of Physics, Stockholm University, SE-10691 Stockholm, Sweden}
\author{I.~Al~Samarai}
\affiliation{D\'epartement de physique nucl\'eaire et corpusculaire, Universit\'e de Gen\`eve, CH-1211 Gen\`eve, Switzerland}
\author{D.~Altmann}
\affiliation{Erlangen Centre for Astroparticle Physics, Friedrich-Alexander-Universit\"at Erlangen-N\"urnberg, D-91058 Erlangen, Germany}
\author{K.~Andeen}
\affiliation{Department of Physics, Marquette University, Milwaukee, WI, 53201, USA}
\author{T.~Anderson}
\affiliation{Dept.~of Physics, Pennsylvania State University, University Park, PA 16802, USA}
% ...

\collaboration{IceCube Collaboration}
\noaffiliation

\date{\today}

\begin{abstract}
As atmospheric neutrinos propagate through the Earth, vacuum-like oscillations are modified by Standard-Model neutral- and charged-current interactions with electrons. Theories beyond the Standard Model introduce heavy, TeV-scale bosons that can produce nonstandard neutrino interactions. These additional interactions may modify the Standard Model matter effect producing a measurable deviation from the prediction for atmospheric neutrino oscillations. 
The result described in this paper constrains nonstandard interaction parameters, building upon a previous analysis of atmospheric muon-neutrino disappearance with three years of IceCube-DeepCore data. The best fit for the muon to tau flavor changing term is $\epsilon_{\mu \tau}=-0.0005$, with a 90\% C.L. allowed range of $-0.0067 <\epsilon_{\mu \tau}< 0.0081$. This result is more restrictive than recent limits from other experiments for $\epsilon_{\mu \tau}$. Furthermore, our result is complementary to a recent constraint on $\epsilon_{\mu \tau}$ using another publicly available IceCube high-energy event selection. Together, they constitute the world's best limits on nonstandard interactions in the $\mu-\tau$ sector.
\end{abstract}

\maketitle

\section{Introduction}

Neutrino flavor change has been observed and confirmed by a plethora of experiments involving solar, atmospheric, reactor, and accelerator-made neutrinos; see~\cite{Esteban:2016qun,Capozzi:2016rtj,Olive:2016xmw} and references therein. This phenomenon, also known as neutrino oscillations due to its periodic behavior, implies that at least two of the Standard Model (SM) neutrinos have a nonzero mass, making this the first established deviation from the SM. The massive three-neutrino model has been very successful in explaining the neutrino data with two mass differences, known as the solar squared-mass difference ($\Delta m^2_{21} \approx 7.5 \times 10^{-5} {\rm eV}^2 $) and the atmospheric squared-mass difference ($|\Delta m^2_{23}| \approx 2.5 \times 10^{-3} {\rm eV}^2 $)~\cite{Esteban:2016qun,Capozzi:2016rtj}.
This information, along with the fact that experiments pursuing direct neutrino mass measurements have yielded only upper limits~\cite{Olive:2016xmw}, leads to the conclusion that neutrinos have masses that are at least six orders of magnitude smaller than those of the charged leptons. Whether these small masses are generated also by the Higgs mechanism, implying the existence of non-interacting right-handed states, or by a different yet-unknown mechanism remains an open question.

Many extensions to the SM that incorporate small neutrino masses have been put forward. A subset that addresses small neutrino masses and, at the same time, unifies the electroweak and strong forces is called ``Grand Unified Theories'' (GUTs).   Some of these GUT models predict the existence of heavy TeV-scale bosons~\cite{King:2017guk}. Searches for direct evidence of these particles have been performed by experiments at the Large Hadron Collider. To date, no evidence has been observed~\cite{Aad:2015owa,Khachatryan:2014hpa}. In this paper, we address these predictions through a complementary search in the neutrino sector, seeking evidence for new flavor-changing neutrino interactions produced by TeV-scale bosons~\cite{Weinberg:1979sa,Fornengo:2001pm,GonzalezGarcia:2004wg,Kopp:2007ne,Antusch:2008tz,Gago:2010}. 

Nonstandard interactions (NSIs) will introduce modifications of the SM potential, which is relevant for matter effects in neutrino flavor oscillations. The effect of the NSI is expected to grow with distance travelled through matter and becomes more relevant as the neutrino energy increases. As a result, the flux of atmospheric neutrinos detected by the IceCube Neutrino Observatory at the South Pole is ideal for such a study~\cite{GonzalezGarcia:2004wg,Esmaili:2013fva}. In the analysis presented here, we use the data set from~\cite{Aartsen:2014yll}, which contains multi-GeV atmospheric neutrinos that traverse large fractions of the Earth before reaching the IceCube detector. Because the neutrino production is predominantly from pion and kaon decays, the neutrino flux has well-understood initial flavor ratios~\cite{PhysRevD.92.023004,PhysRevD.83.123001}.

Current bounds on NSI are reported in~\cite{Biggio:2009nt,Biggio:2009kv,Escrihuela:2009up}, and current reviews are given in~\cite{Ohlsson:2012kf,Miranda:2015dra,Davidson:2003ha,Escrihuela:2011cf}. In fact, independent studies of high-energy atmospheric neutrinos using public IceCube data~\cite{Salvado:2016uqu} as well as studies with public Super-Kamiokande data~\cite{Fukasawa:2015jaa} have already been performed, obtaining strong constraints on NSI parameters. Regarding the latter, the Super-Kamiokande collaboration has also performed an analysis on NSI parameters~\cite{Mitsuka:2011ty}. The IceCube studies have so far only used high-energy public data sets, but no low-energy sets. This motivates the presented search,
where we focus on the NSI parameter $\epsilon_{\mu \tau}$, which modifies the $\nu_\mu \rightarrow \nu_\tau$ flavor transition.

The rest of this paper is structured as follows. In section~\ref{sec:matter}, we review neutrino oscillations in matter. In section~\ref{sec:nsi}, we describe the NSI theory used in this work. Then in section~\ref{sec:detector}, we describe the IceCube experiment, and in~\ref{sec:systematics} we discuss the main systematics of this analysis. Section~\ref{sec:result} contains the main results of this paper, and in section \ref{sec:conclusions} we conclude.

\section{Matter Effects in Neutrino Oscillations\label{sec:matter}}

Neutrinos are produced in flavor eigenstates, but travel as mass eigenstates, meaning that a certain flavor of neutrino produced at the source may later interact as a different flavor~\cite{Pontecorvo:1967fh,Maki:1962mu}.  At its simplest, when neutrinos travel through vacuum, the oscillation length is given by $L_{osc} = 2.5~{\rm km} \left( E /{\rm GeV} \right) \left( \Delta m^2   /  {\rm eV^2} \right)^{-1}$~\cite{Olive:2016xmw}.

Since neutrinos interact via neutral- and charged-current weak interactions, neutrino oscillations are modified as matter is traversed. In particular, the propagating neutrino -- which is a mixture of electron, muon, and tau flavors -- will experience a flavor-dependent matter potential. The relevant potential difference is produced by charged-current coherent forward scattering from electrons in the Earth. We will refer to this as ``matter effect,'' and it is closely related to the Mikheyev-Smirnov-Wolfenstein (MSW) effect~\cite{Wolfenstein:1977ue,Mikheev:1986gs} observed in solar neutrino experiments~\cite{Abe:2010hy,Ahmad:2001an,Aharmim:2011vm,Bellini:2011rx}. Indications of matter effects~\cite{Freund:1999gy,Akhmedov:1998xq} in Earth-based oscillation experiments can be extracted from global fits to long-baseline and atmospheric neutrino data sets~\cite{Gonzalez-Garcia:2013usa}. 

\section{Nonstandard Neutrino Interactions\label{sec:nsi}}

Nonstandard neutrino interactions can be modeled as an additional term in the neutrino Hamiltonian, similar to the conventional matter potential term. The latter effect is included in the neutrino Hamiltonian as a single potential, $V_{CC}$, which modifies the flavor transition probabilities. The potential V$_{CC}$ is proportional to the Fermi coupling constant $G_f$ and the density of electrons $n_e$, i.e., $V_{CC} = \sqrt{2} G_{f} n_e$.

Adding interactions with nonstandard bosons to the Hamiltonian takes a similar form, but with additional components. To consider all possible flavor-violating interactions, a term $\epsilon_{\alpha\beta}$ ($\alpha,\beta = e,\mu,\tau$) scales all possible flavor-violating and conserving contributions. For definiteness, in this analysis, we consider nonstandard interactions between neutrinos and down quarks (other assumptions such as for up quarks can be approximated by rescaling our results).  For this reason, a factor of $n_d=3n_e$(to account for the fact that down-quarks are approximately three times as abundant as electrons in the Earth) was used instead of $n_e$ as in the case of the SM matter effect. The total Hamiltonian is then
\begin{equation}
H_{3\nu} = \frac{1}{2E_{\nu}} U M^2 U^{\dag}+V_{CC}diag(1,0,0) + V_{CC}\frac{n_d}{n_e} \epsilon,
\end{equation}
where $E_\nu$ is the neutrino energy, $U$ is the neutral lepton mixing matrix (also known as the Pontecorvo-Maki-Nakagawa-Sakata matrix~\cite{Pontecorvo:1967fh,Maki:1962mu}), $M^2$ is a diagonal matrix containing the square-mass differences, and
\begin{equation}
\epsilon = 
 \left( \begin{array}{ccc}
\epsilon_{ee} & \epsilon_{e\mu} & \epsilon_{e\tau} \\
\epsilon^*_{e\mu} & {\epsilon_{\mu \mu}} &{\epsilon_{\mu \tau}} \\
\epsilon^*_{e\tau} & {\epsilon^*_{\mu \tau}} & {\epsilon_{\tau \tau}} \\
\end{array} \right),
\end{equation}
the NSI strength matrix. Accordingly, the addition of the NSI terms amounts to introducing six additional effective parameters if one accounts for hermicity, unitarity constraints, and the possibility of making the Hamiltonian traceless without loss of generality; see~\cite{Liao:2016hsa}. However, for experiments like Super-Kamiokande and IceCube, the terms that correspond to $\nu_{\mu}$ or $\nu_{\tau}$ interactions will dominate. This is because the atmospheric neutrino flux in the GeV energy range is dominated by $\nu_\mu$, which primarily transform into $\nu_\tau$ as they travel through the Earth. \cite{Akhmedov:1999uz,Fukugita:2003en}.

\begin{figure}[t]
\begin{center}
\includegraphics[width=\columnwidth]{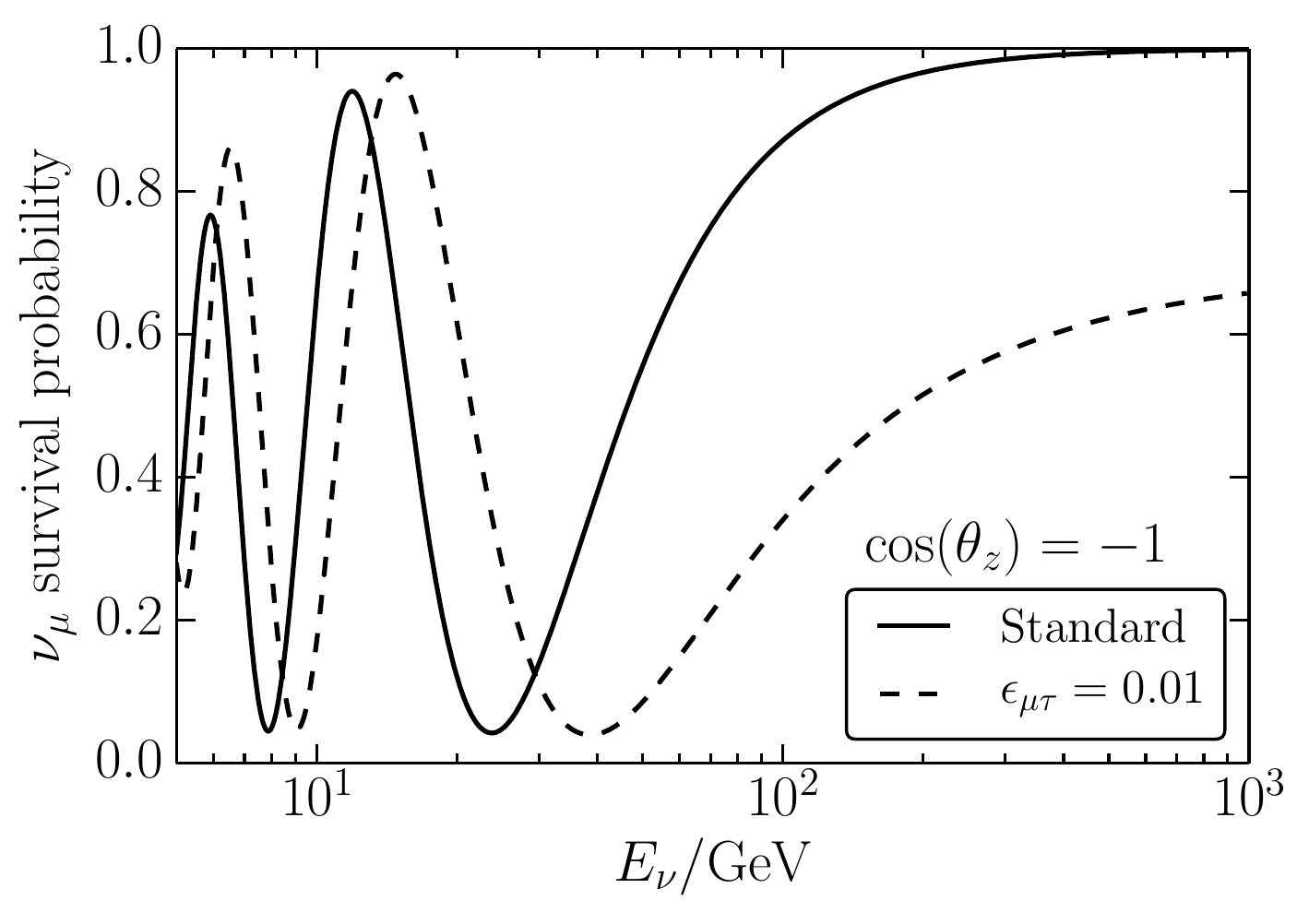}
\includegraphics[width=\columnwidth]{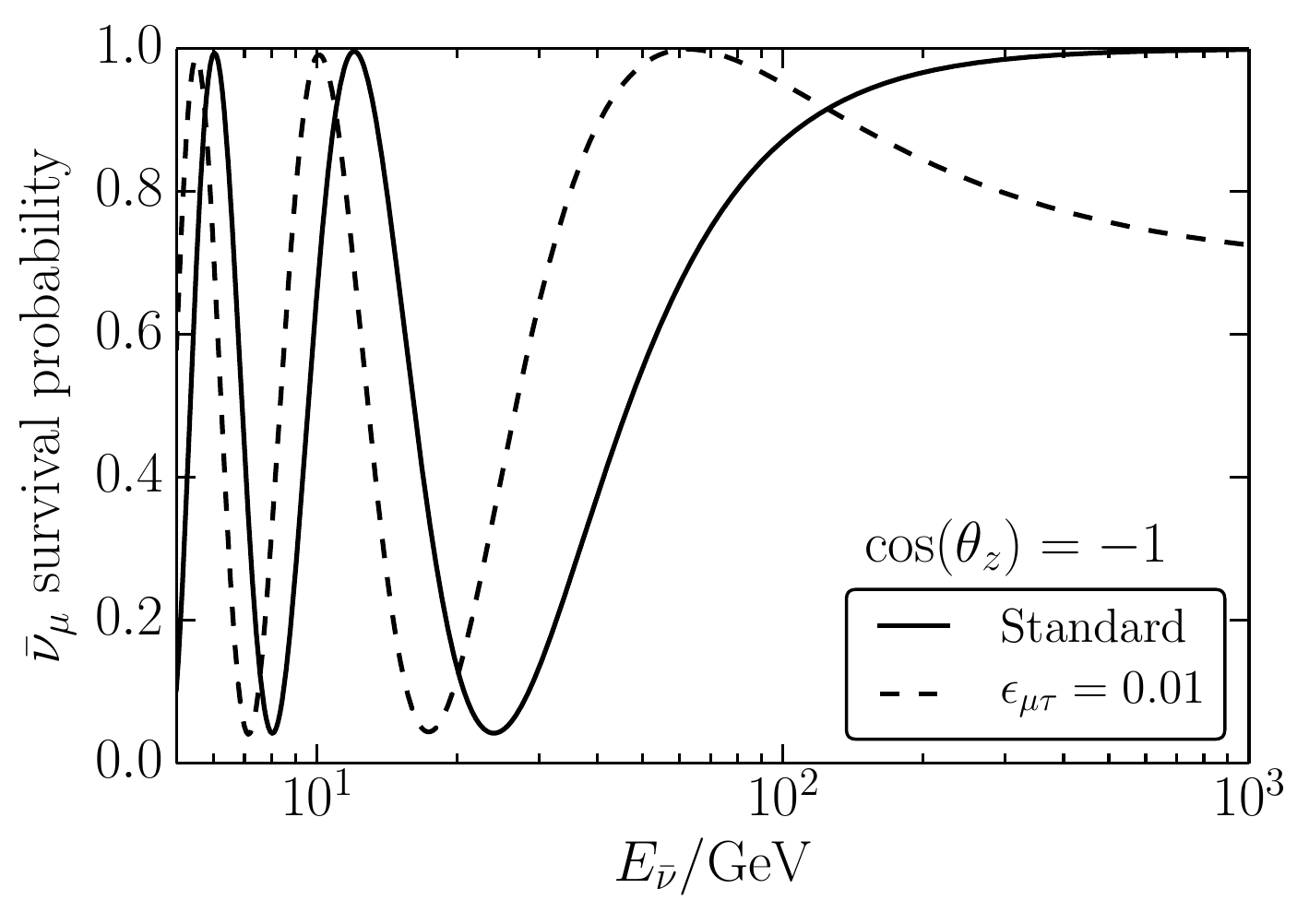}
\end{center}
\caption{\label{fig:nsioscplots} Muon neutrino (top) and antineutrino (bottom) survival probability at zenith angle $\cos\theta=-1$, corresponding to vertically up going neutrinos that traverse the entire diameter of the Earth, for global best-fit oscillations (solid) and $\epsilon_{\mu \tau}=0.01$ NSI, close to the current Super-Kamiokande limits (dashed)\cite{Mitsuka:2011ty}.
NSI effects are visible in the full neutrino energy range of 10-1000 GeV.}
\end{figure}

SM matter effects and NSI can be distinguished using the energy and arrival direction distributions of observed flavor-violating transitions. The neutrino flavor oscillations due to the well-established mass differences have been observed from atmospheric neutrinos predominately at energies initially below 10 GeV~\cite{Fukuda:1998mi} and recently up to 56 GeV~\cite{Aartsen:2014yll}. The observation of atmospheric neutrino oscillations at two different energy ranges but at the same ratio of baseline to energy ($L/E$) tests the massive three neutrino paradigm and highlights the complementarity of neutrino experiments at different energy ranges. In contrast, the signal predicted for the dominant muon-neutrino to tau-neutrino NSI, parametrized by the coupling $\epsilon_{\mu \tau}$, has a smaller magnitude but can be seen over a larger range of energies, as shown in Fig.~\ref{fig:nsioscplots}. Therefore, the optimal method for searching for an NSI signal due to $\epsilon_{\mu \tau}$ is to use a large range of neutrino energies, where one expects a combined effect of the NSI and oscillations in the low-energy region and an exclusively  NSI signal in the high-energy region. In particular, we note that IceCube's range extends to higher energies than that of previous studies, thus giving us greater sensitivity.

A study by Super-Kamiokande~\cite{Mitsuka:2011ty}, using a two-neutrino approximation, focused on the NSI parameters $\epsilon^\prime = \epsilon_{\tau \tau}-\epsilon_{\mu \mu}$ and $\epsilon_{\mu \tau}$. Prior to works using IceCube data, this resulted in the world's best limit with $\vline \epsilon_{\mu \tau} \vline < 0.011 $ at 90\% CL.
As in the Super-Kamiokande study, we choose to only consider the dominant NSI terms, so the $\nu_e$ terms are set to zero, and the hermiticity of $\epsilon$ is also assumed. Thus, the NSI sector reduces to a two by two matrix, so the CP-violating phase can be rephased, i.e., we assume $\epsilon_{\mu \tau}$ to be real. As can be seen in~\cite{Miranda:2015dra}, the neutrino mass ordering is degenerate with the sign of $\epsilon_{\mu\tau}$, and the muon neutrino survival probability is symmetric under sign change of $\epsilon^\prime$. Given that $\epsilon^\prime$ is highly correlated with $\epsilon_{\mu \tau}$ in this analysis, we set $\epsilon^\prime$ to zero. Also, for definiteness, we assume normal ordering. Note that these degeneracies restrict the interpretation of our results~\cite{Mitsuka:2011ty,GonzalezGarcia:2011my,Miranda:2015dra,Coloma:2016gei}.

\section{The IceCube Detector\label{sec:detector}}

\begin{figure}
\centering
\includegraphics[width=\columnwidth]{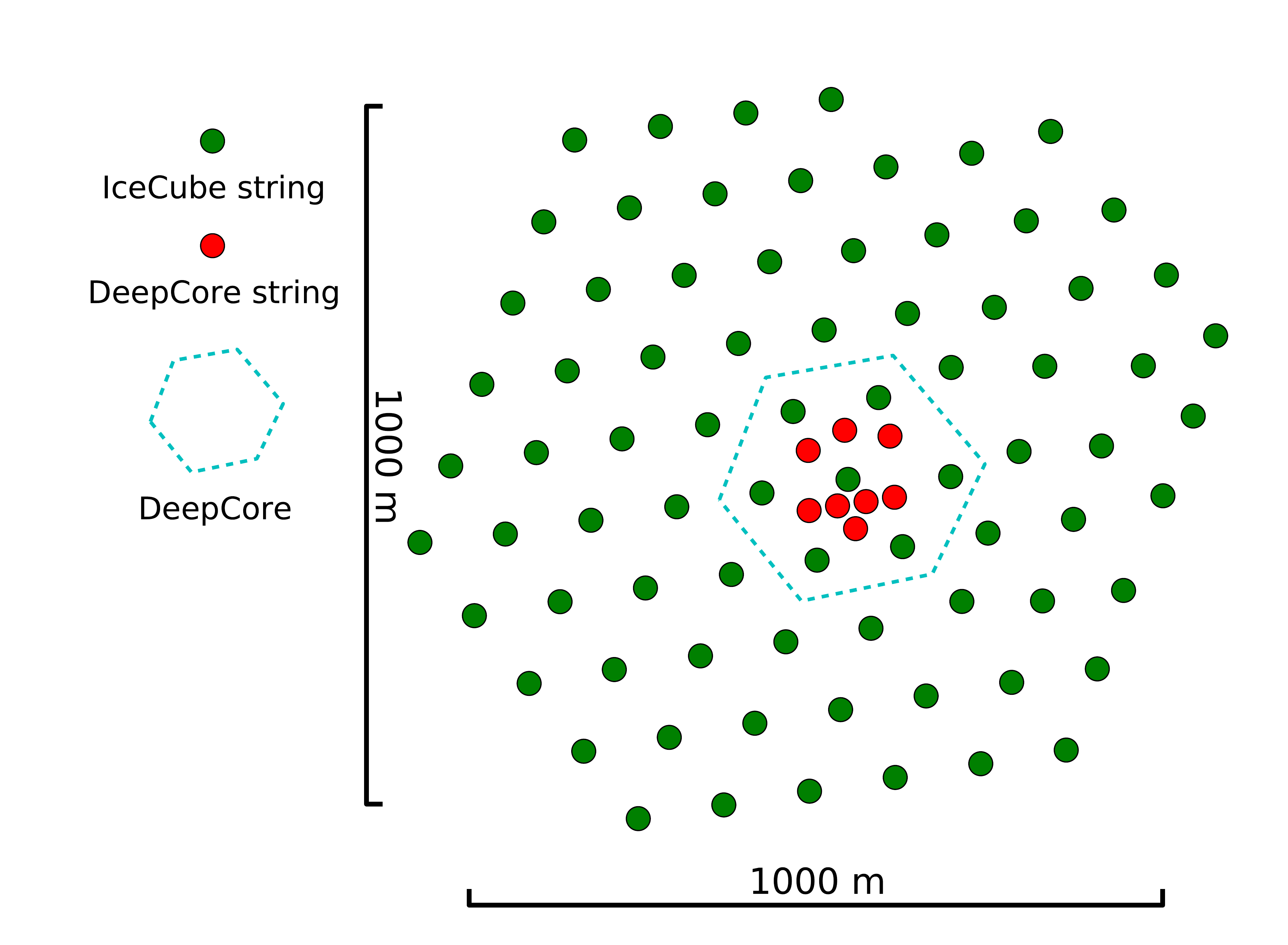} 
\caption{\label{fig:detector} Detector geometry: green circles represent IceCube strings and red ones deep core strings.}
\end{figure}

IceCube is a 1 km$^3$ neutrino detector~\cite{Achterberg:2006md,Abbasi:2008aa,Aartsen:2016nxy} embedded in the ice at the South Pole; see Fig.~\ref{fig:detector}. The detector consists of 86 strings, each with 60 10-inch photomultiplier tubes enclosed in glass spheres, called Digital Optical Modules (DOMs). Of those strings, 78 are separated by a distance of approximately 125 m, with DOMs on each string separated by 17~m. An additional infill extension, DeepCore~\cite{Abbasi2012615}, consists of 8 strings separated by about 75~m, with DOMs on each string separated by 7~m. Secondary particles produced when neutrinos interact in the ice, induce Cherenkov radiation, which is then detected by the DOMs. Muons produce distinctively long tracks. This topology can be reconstructed to determine the angle of the muon with a resolution of $12^{\circ}$ at 10~GeV, improving to $6^{\circ}$ at 40~GeV~\cite{Aartsen:2014yll}. The energy of the muon can be measured from its track length, while the energy of the hadronic shower produced in the neutrino interaction can be estimated from the total amount of light in the detector. Thus, the muon energy, estimated from the track length, added to the reconstructed shower energy is a proxy for the neutrino energy. The closely spaced DOMs of the DeepCore extension allow measuring the neutrino energy down to neutrino energies of about 5~GeV, with a median resolution of 30\% at 8~GeV, which improves to 20\% at 20~GeV~\cite{Aartsen:2014yll}. This analysis makes use of neutrinos that reach the detector from below the Earth's horizon. This serves two purposes: first it greatly diminishes atmospheric muon contamination and, second, it allows for large matter effects.

\section{IceCube sensitivity to nonstandard interactions and  systematic uncertainties\label{sec:systematics}}

\subsection{Sensitivity and data set}

\begin{figure*}
\centering
\includegraphics[width=0.90\columnwidth]{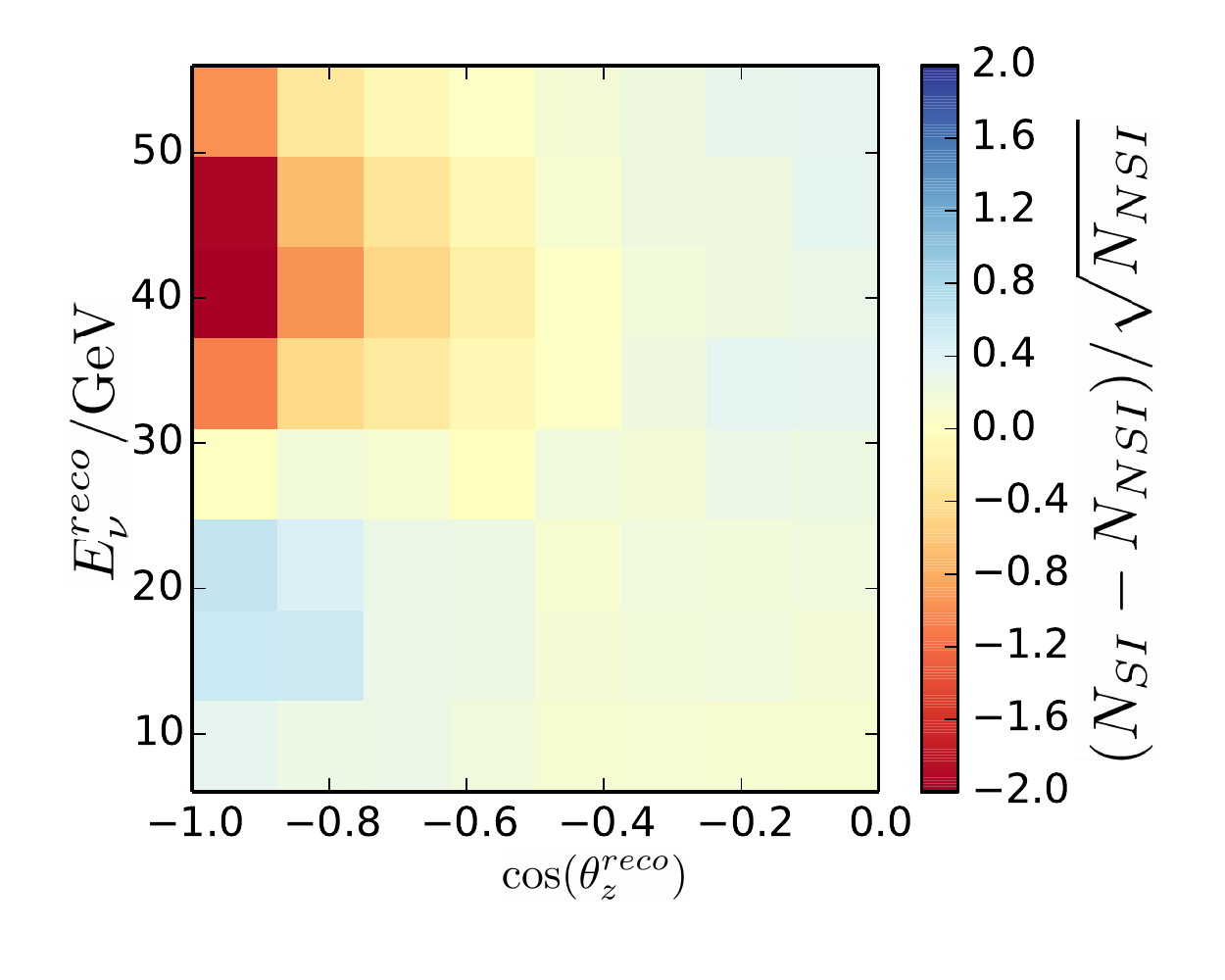} 
\includegraphics[width=0.90\columnwidth]{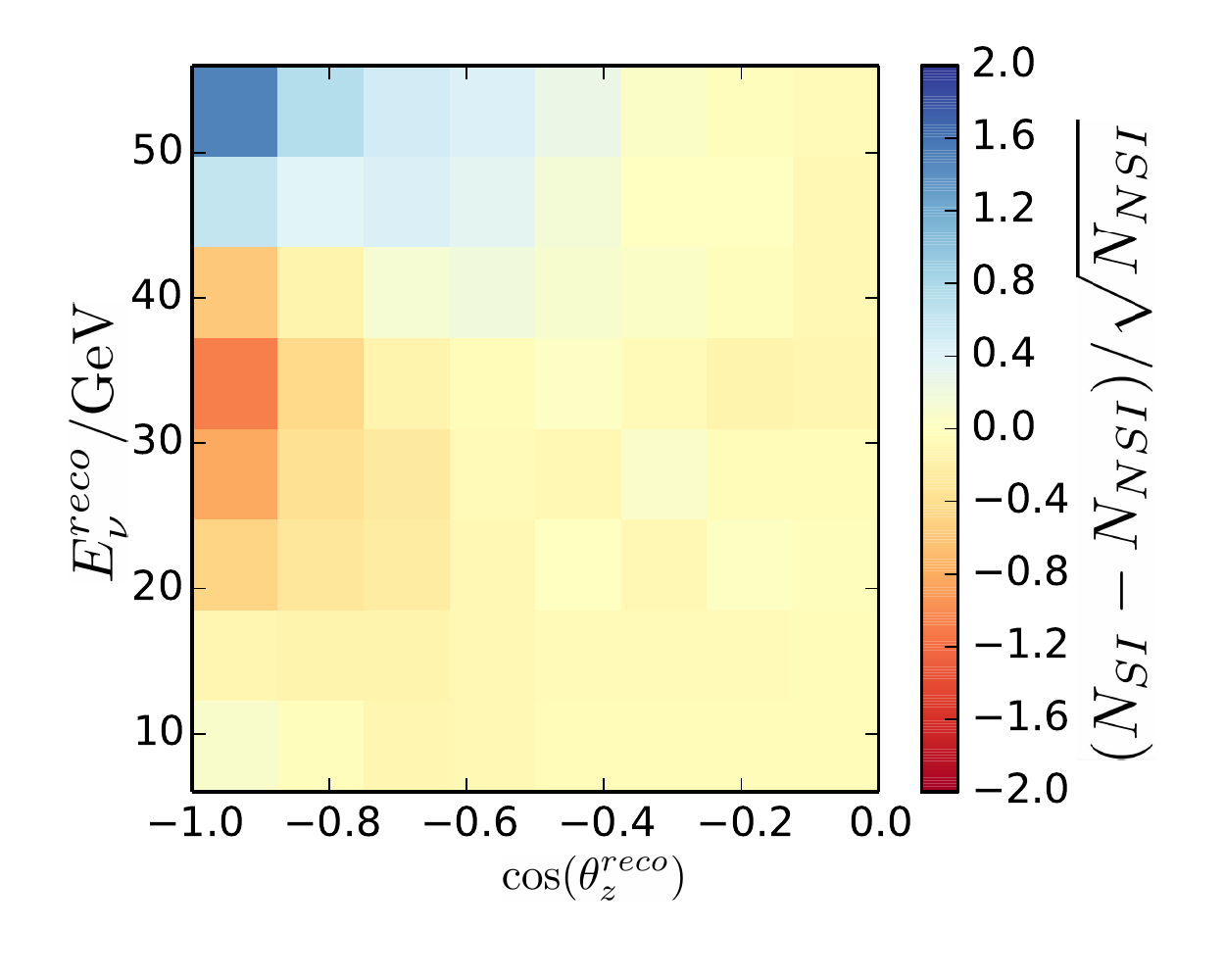}
\caption{\label{fig:sigcalc} Expected pulls of predicted event numbers as a function of neutrino energy and zenith angle. The left (right) panel compares $\epsilon_{\mu \tau}$=-0.01 ($\epsilon_{\mu \tau}$=0.01) to the standard neutrino oscillation matter effects (SI) expectation.}
\end{figure*}

IceCube has measured neutrino oscillation parameters by searching for a deficit of neutrinos traveling through Earth and interacting in the detector.  In IceCube, the $\nu_\mu$ disappearance probability peaks at $\sim 25~{\rm GeV}$ for straight up going events, but the oscillation signal is measurable up to about $100~{\rm GeV}$, as shown in Fig.~\ref{fig:nsioscplots}. In 2014, IceCube published the result of fitting 5174 events from three years of data taken with the complete IceCube detector, obtaining three-neutrino oscillation parameters to a precision comparable with that from dedicated neutrino oscillation experiments~\cite{Aartsen:2014yll}. This study uses a three-neutrino formalism of the neutrino survival probabilities to calculate limits on the $\epsilon_{\mu \tau}$ parameter.  We use  the publicly available nuSQuIDS neutrino survival probability package~\cite{Delgado:2014kpa,nusquids}, which has a robust implementation of NSI and uses a detailed Earth density profile~\cite{Dziewonski:1981xy}.
Simulated events are weighted with the Honda et al. atmospheric neutrino model~\cite{PhysRevD.92.023004}, then are binned in an 8$\times$8 matrix in reconstructed energy, from 6.3~GeV to 56.2~GeV, and zenith angle, from $\cos\theta_z^{reco}=-1$ to $\cos\theta_z^{reco}=0$ (see Fig.~\ref{fig:sigcalc}). To determine the expected sensitivity for values of $\epsilon_{\mu \tau}$ in the range of the Super-Kamiokande limit, the total number of events expected with and without NSI effects were calculated as shown in Fig.~\ref{fig:sigcalc}.

\subsection{Systematic uncertainties}

Systematic uncertainties that we have included as nuisance parameters in the fit are:
\begin{description}
\item[{\it Oscillation parameters}] simultaneously fit for the standard oscillation parameters $\sin^2(\theta_{23})$ and $\Delta m^2_{23}$ as nuisance parameters.
\item[{\it Ice column scattering coefficient}] scattering of light in the ice that formed within the hole after the DOMs were inserted~\cite{Aartsen:2013rt}.  This ice contains bubbles  that are not found in the bulk ice of the detector. The latter is well studied using flashers and well modeled. The additional bubbles increase the scattering of light, affecting the effective angular efficiency of our DOMs; see~\cite{Aartsen:2013rt} for details.
\item[{\it Optical efficiency}] the uncertainty in the photon response of the optical modules due to many effects, including photocathode response and obscured regions due to cabling.
\item[{\it Overall normalization ($N$)}] parameter that scales the event rate expectation freely. This absorb overall normalization uncertainties due to absolute DOM efficiencie and total cosmic ray flux.
\item[{\it Relative $\nu_e$ to $\nu_\mu$ normalization ($N_{e/\mu}$)}] relative normalization of the electron neutrinos to atmospheric muon neutrinos. 
\item[{\it Atmospheric muon fraction ($R_\mu$)}] normalization of cosmic ray muons that pass the cuts. The distribution of this background was obtained using a data driven method~\cite{Aartsen:2014yll}. 
\end{description}

\begin{figure*}
\centering
\includegraphics[width=\textwidth]{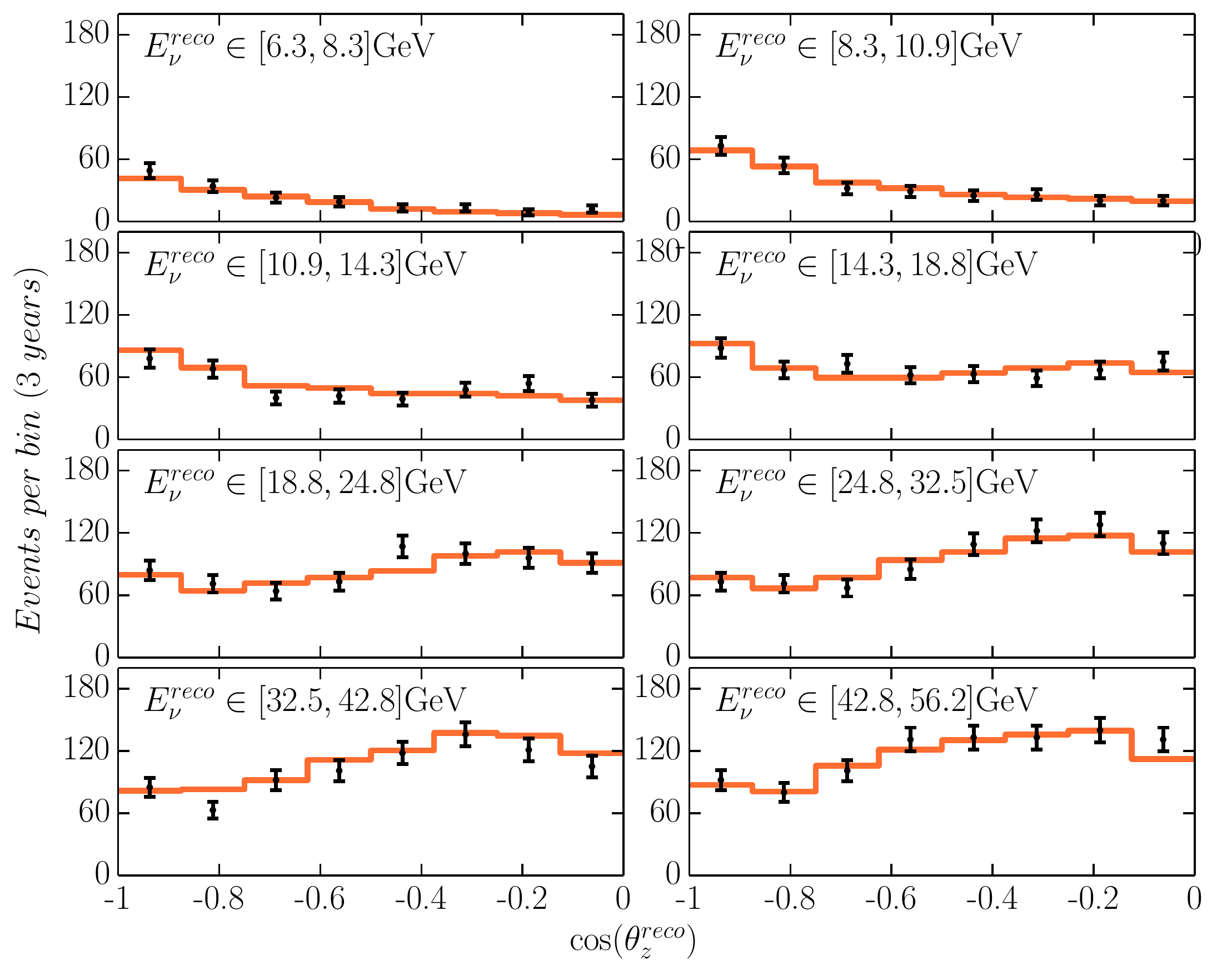}
\caption{\label{fig:datamc} Data (black points) and Monte Carlo (solid orange line) comparisons for this analysis, after the fiducial volume cut and fit of the nuisance parameters, as a function of the arrival direction, $\cos(\theta_{z}^{reco})$, in the eight different energy bins.}
\end{figure*}

\FloatBarrier

\begin{description}
\item[{\it Spectral index ($\gamma$)}] the exponent describing the energy dependence of the incoming cosmic ray spectrum. This systematic in part accounts for uncertainties due to hadronization processes~\cite{Katori:2014fxa}.
\end{description}
For a more detailed discussion of these systematic effects, see~\cite{Aartsen:2017bap}.

\section{Result\label{sec:result}}

\begin{flushleft}
\begin{table}[t!]
\centering      
\begin{tabularx}{\columnwidth}{l c c c}  
\hline\hline                        
Parameter & Prior Center & Prior Width & Fit Value \\ [0.5ex] 
\hline                    
$\epsilon_{\mu\tau}$ & - & - & -0.0005  \\  
$\sin^2(\theta_{23})$ & - & - & 0.52  \\  
$\Delta m^2_{31}/{\rm eV^2}$ & - & - & $2.62\times 10^{-3}$   \\ 
Ice column scat. (cm$^{-1}$) & 0.02 & 0.01  & 0.02  \\ 
Optical efficiency ($\%$) & 100 & 10 & 101  \\ 
Overall norm.. ($N$) & - & - & 1.00  \\
Rel. $\nu_e$ norm. ($N_{e/\mu}$) & 1 & 0.20 & 1.03  \\
Atmospheric $\mu$ ($R_\mu$) & - & - & 0.0  \\ 
Spectral index ($\gamma$) & 2.70 & 0.05 & 2.67\\
[1ex]       
\hline     
\end{tabularx} 
\caption{List of the best-fit parameters obtained in this analysis. When priors are listed, they are Gaussian, and the width corresponds to the one sigma range. Values obtained at the best-fit point are also listed.}
\label{table:systematics}  
\end{table}
\end{flushleft}

In order to constrain the NSI parameter $\epsilon_{\mu \tau}$, we employ the same data set and event selection in this analysis as was used in~\cite{Aartsen:2014yll}. This analysis has the same energy, zenith angle resolution, and systematic uncertainties as the analyses in~\cite{Aartsen:2014yll,Aartsen:2017bap} with an additional fiducial volume cut, resulting in a  final sample to 4625 events~\cite{Aartsen:2017bap}. The data and Monte Carlo are in good agreement after the fit, as shown in Fig.~\ref{fig:datamc}.

To determine the best-fit oscillation parameters, the simulated data distributions are compared to the data bin-by-bin. Minimizing the Poisson likelihood value of the data given the Monte Carlo, modified by the nuisance parameters (as described in~\cite{Aartsen:2014yll,Aartsen:2017bap}), determines the final best-fit parameters. The 90\% confidence level limits are then calculated using the difference from the best-fit likelihood, assuming Wilks' theorem applies~\cite{wilks1938}. To make the comparison to~\cite{Salvado:2016uqu}, we also calculate the credibility regions by integrating the profiled likelihood using a uniform prior on $\epsilon_{\mu\tau}$ and profiling over the nuisance parameters. This procedure is found to be in good agreement with the result obtained using Wilk's theorem.

\begin{figure}[h!]
\centering
\includegraphics[width=\columnwidth]{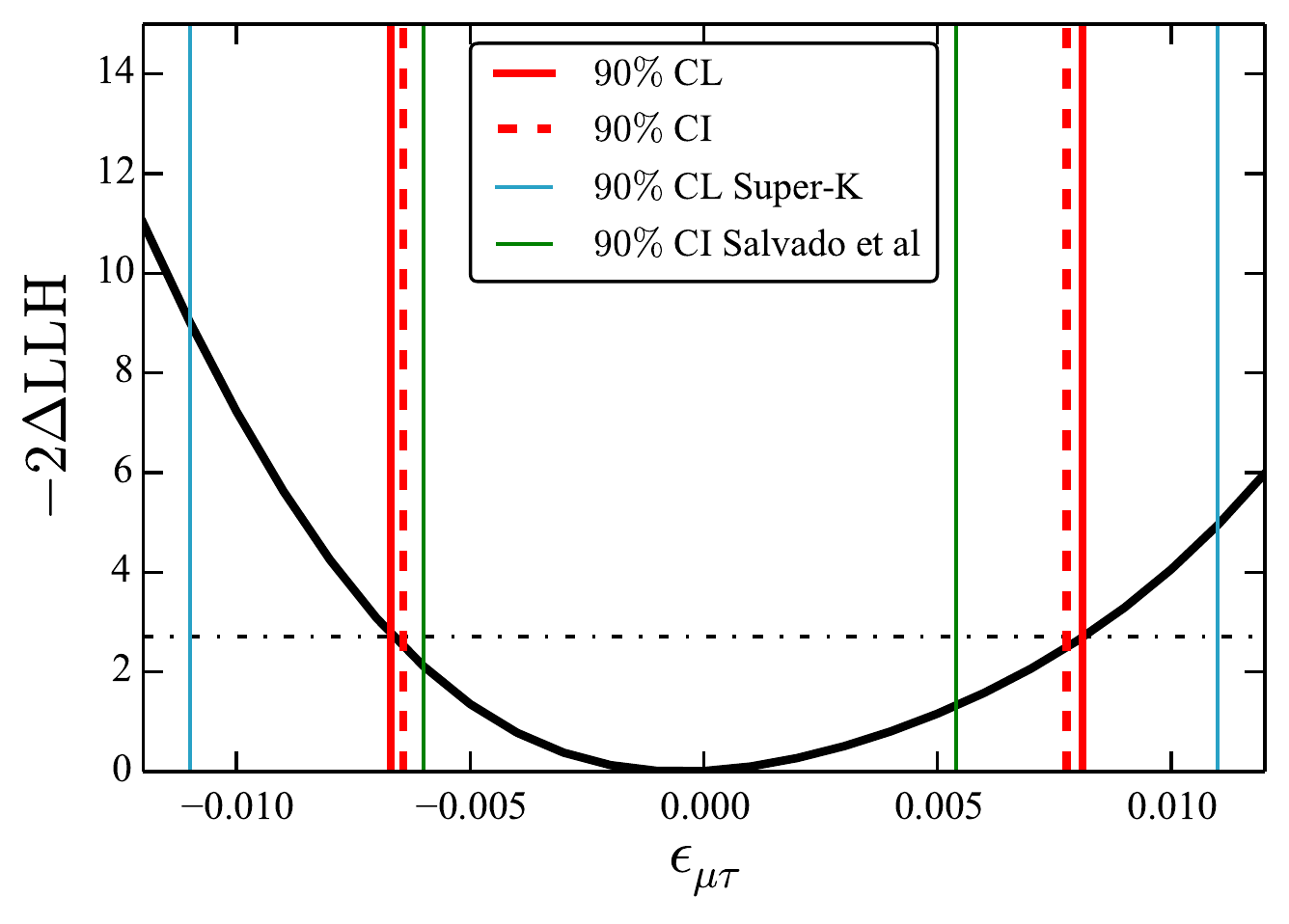}
\caption{\label{fig:NSIFit} Confidence limits from this analysis on the NSI parameter $\epsilon_{\mu \tau}$ using the event selection from~\cite{Aartsen:2014yll,Aartsen:2017bap} shown as solid vertical red lines.  Similarly, dashed vertical red lines show the 90\% credibility interval using a flat prior on $\epsilon_{\mu\tau}$ and where we have profiled over the nuisance parameters. The light blue vertical lines show the Super-Kamiokande 90$\%$ confidence limit~\cite{Mitsuka:2011ty}. The light green lines show the 90\% credibility region from~\cite{Salvado:2016uqu}. Finally, the horizontal dash-dot line indicates the value of $-2\Delta {\rm LLH}$ that corresponds to a 90$\%$ confidence interval according to Wilks' theorem.}
\end{figure}

\begin{figure}[h!]
\centering
\includegraphics[width=\columnwidth]{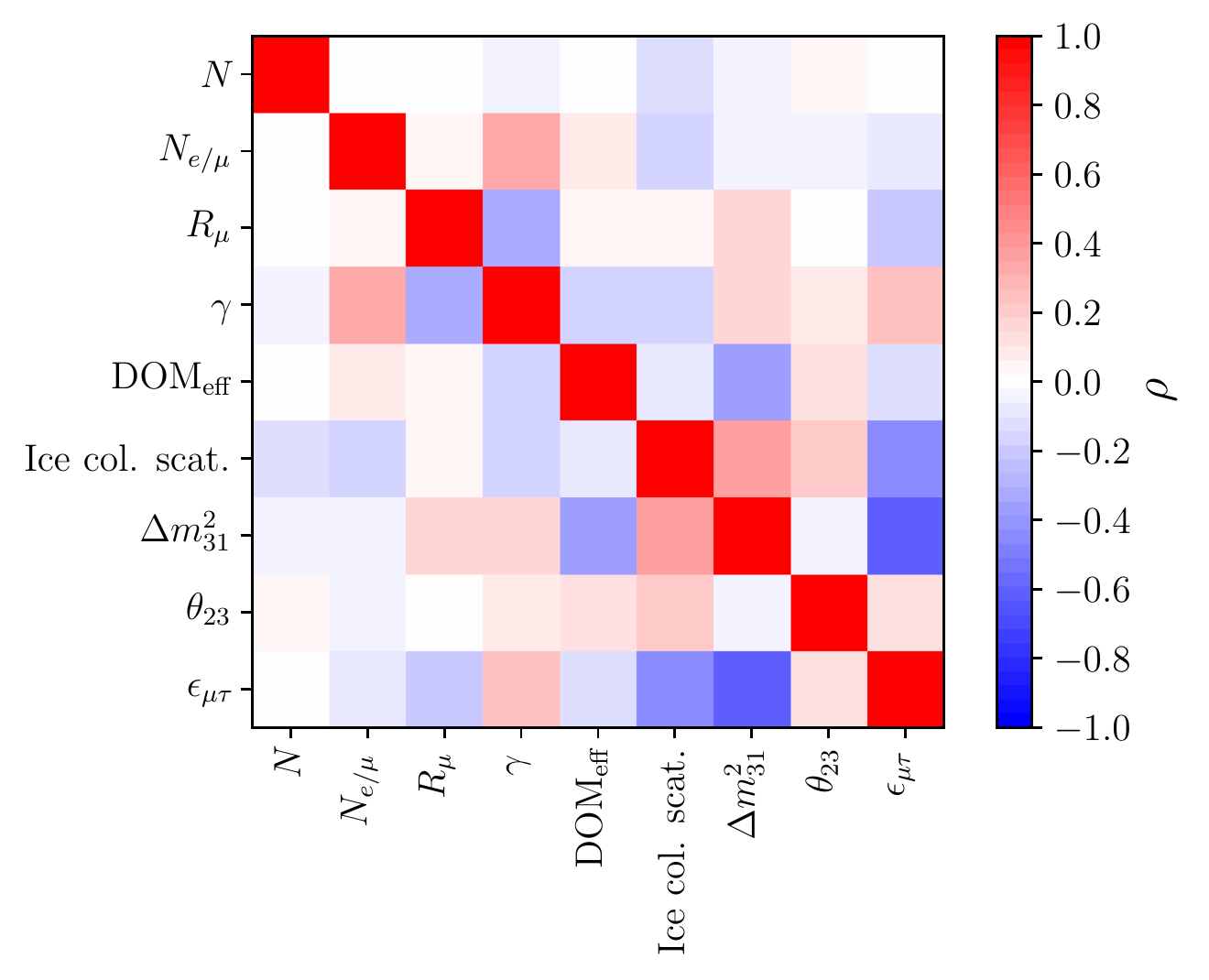}
\caption{\label{fig:Correlation} Correlation matrix of the nuisance and physics parameters considered in this analysis calculated at the maximum likelihood solution. The color scale show the correlation coefficient ($\rho$).}
\end{figure}

The resulting constraints on the NSI parameters are shown in Fig.~\ref{fig:NSIFit}, with the best-fit values for the systematic parameters shown in Table \ref{table:systematics}. Priors on the atmospheric and detector nuisance parameters are the same as in~\cite{Aartsen:2014yll}. Furthermore, Fig.~\ref{fig:Correlation} shows the correlation between the fit parameters at the best fit of oscillation and nuisance parameters. The mass-squared difference $\Delta m^2_{31}$ exhibits the strongest correlation with $\epsilon_{\mu\tau}$. This is to be expected from existing correlations and degeneracies in the oscillation probability~\cite{Liao:2016hsa}.
Finally, the change in the oscillation parameters compared to~\cite{Aartsen:2014yll} have been demonstrated to be caused by the additional cut on the fiducial volume.

For this analysis, the best fit is at $\epsilon_{\mu \tau}=-0.0005$. The 90\% C.L. range is $-0.0067<\epsilon_{\mu \tau}< 0.0081$. This result is consistent with the Super-Kamiokande limits for $\epsilon_{\mu \tau}$~\cite{Mitsuka:2011ty} and represents an independent determination of the parameter. To compare with this, in Fig.~\ref{fig:NSIFit} we show the results from~\cite{Salvado:2016uqu} obtained using public IceCube high-energy data. Fig.~\ref{fig:nsioscplots} shows that the signal for $\epsilon_{\mu \tau}$ is largest in the region above 100 GeV. A planned extension of this study including a sample of events above 100 GeV would significantly improve constraints on NSI parameters~\cite{Esmaili:2013fva}.

\section{Conclusions\label{sec:conclusions}}

The existence of physics beyond the Standard Model has been suggested by the nonzero neutrino mass, in addition to the existence of dark matter. Extensions of the Standard Model that explain these observations could lead to a modified strength of neutrino interactions in standard matter.
Experiments like IceCube have the potential to constrain these nonstandard interactions with greater precision than previous experiments.

Our best fit of the NSI flavor-changing parameter yields $\epsilon_{\mu \tau}=-0.0005$, with a 90$\%$ C.L. range of $-0.0067 < \epsilon_{\mu \tau} < 0.0081$. This result is comparable to with a slight improvement over the Super-Kamiokande limits for $\epsilon_{\mu \tau}$ ($|\epsilon_{\mu \tau}|<0.011$ at 90\% C.L.). 
A recent study~\cite{Salvado:2016uqu} using IceCube public data obtained constraints which are slightly better than the ones shown in this paper. These constraints are also shown in Fig.~\ref{fig:NSIFit} and are complementary to our result as they are affected by different systematics and make use of a different energy regime.

\begin{acknowledgments}

We acknowledge the support from the following agencies:
U.S. National Science Foundation-Office of Polar Programs,
U.S. National Science Foundation-Physics Division,
University of Wisconsin Alumni Research Foundation,
the Grid Laboratory Of Wisconsin (GLOW) grid infrastructure at the University of Wisconsin - Madison, the Open Science Grid (OSG) grid infrastructure;
U.S. Department of Energy, and National Energy Research Scientific Computing Center,
the Louisiana Optical Network Initiative (LONI) grid computing resources;
Natural Sciences and Engineering Research Council of Canada,
WestGrid and Compute/Calcul Canada;
Swedish Research Council,
Swedish Polar Research Secretariat,
Swedish National Infrastructure for Computing (SNIC),
and Knut and Alice Wallenberg Foundation, Sweden;
German Ministry for Education and Research (BMBF),
Deutsche Forschungsgemeinschaft (DFG),
Helmholtz Alliance for Astroparticle Physics (HAP),
Initiative and Networking Fund of the Helmholtz Association,
Germany;
Fund for Scientific Research (FNRS-FWO),
FWO Odysseus programme,
Flanders Institute to encourage scientific and technological research in industry (IWT),
Belgian Federal Science Policy Office (Belspo);
Marsden Fund, New Zealand;
Australian Research Council;
Japan Society for Promotion of Science (JSPS);
the Swiss National Science Foundation (SNSF), Switzerland;
National Research Foundation of Korea (NRF);
Villum Fonden, Danish National Research Foundation (DNRF), Denmark

\end{acknowledgments}

\bibliography{nsibiblio}

\end{document}